\documentclass[twocolumn,secnumarabic,amssymb, nobibnotes, aps, prd]{revtex4-1}

\usepackage{color}

\usepackage{epstopdf}

\usepackage{float,subfig,physics}
\usepackage{graphicx} 
\usepackage[T1]{fontenc}

\newcommand{\Iambda}{%
  \mathnormal{\Lambda}%
}

\newcommand{\angles}[1]{%
	\left \langle #1 \right \rangle%
}

\newcommand{\argmin}{\operatornamewithlimits{arg\,min}}

\newcommand{\subfigimg}[3][,]{%
  \setbox1=\hbox{\includegraphics[#1]{#3}}
  \leavevmode\rlap{\usebox1}
  \rlap{\raisebox{\dimexpr\ht1-1\baselineskip}{\footnotesize(#2)}}
  \phantom{\usebox1}
}

\newcommand*{\citen}[1]{
  \begingroup%
    \romannumeral-`\x
    \setcitestyle{numbers}%
    \cite{#1}%
  \endgroup   
}

\usepackage{chngcntr}%
\newcommand{\beginsupplement}{%
        \appendix
        \setcounter{table}{0}
        \renewcommand{\thetable}{S\arabic{table}}%
        \setcounter{figure}{0}
        \renewcommand{\thefigure}{S\arabic{figure}}%
        \counterwithout{equation}{section}%
        \setcounter{equation}{0}
        \renewcommand{\theequation}{S\arabic{equation}}%
}

\begin{document}


\title{\textbf{Clustering and assembly dynamics of a one-dimensional microphase former}}


\author{{Yi Hu\textit{$^{a}$} and Patrick Charbonneau$^{\ast}$\textit{$^{a,b}$}}\\
{\small \em $^a $Department of Chemistry, Duke University, Durham, North Carolina 27708, USA\\
$^b$Department of Physics, Duke University, Durham, North Carolina 27708, USA\\
$^*$ Email: patrick.charbonneau@duke.edu
}}


\date{\today}

\begin{abstract}
Both ordered and disordered microphases ubiquitously form in suspensions of particles that interact through competing short-range attraction and long-range repulsion (SALR).
While ordered microphases are more appealing materials targets, 
understanding the rich structural and dynamical properties of their disordered counterparts is essential to controlling their mesoscale assembly. Here, we study the disordered regime of a one-dimensional (1D) SALR model, whose simplicity enables detailed analysis by transfer matrices and Monte Carlo simulations. We first characterize the signature of the clustering process on macroscopic observables, and then assess the equilibration dynamics of various simulation algorithms. We notably find that cluster moves markedly accelerate the mixing time, but that event chains are of limited help in the clustering regime. These insights will guide further study of three-dimensional microphase formers.
\end{abstract}


\maketitle


\section{Introduction}
Controlling the mesoscopic assembly of particles with short-range attraction and long-range repulsion (SALR) interactions remains an open challenge for experimental soft matter~\cite{zhuang2016recent}. Although similar interactions in diblock copolymers result in the robust formation of both periodic and disordered microphases\cite{bates1990block,bates1999block,kim2009block}, only the latter have thus far been observed in colloidal suspensions~\cite{sciortino2004equilibrium,stradner2004equilibrium,campbell2005dynamical,klix2010structural,zhang2012non}. From a theoretical viewpoint the situation is better controlled, thanks to various theoretical and methodological advances~\cite{ciach2013origin,broccio2006structural,toledano2009colloidal,ciach2016density,RN116,zhuang2016equilibrium,zhuang2016recent,zhuang2017communication}. But even then the formation of disordered microphases remains only partially understood. Because equilibrating these structures is likely key to ordering periodic microphases, elucidating their assembly is an important hurdle to overcome. 



Using numerical simulations to characterize the disordered microphase regime faces a couple of key difficulties. First, precisely identifying the onset of microphase formation can be challenging. The assembly of disordered mesophases at low colloid density is akin to that of surfactants micelles, as suggested by their shared Landau-Brazovskii free energy functional~\cite{ciach2013origin}. Once the particle concentration exceeds the critical cluster density (ccd)\cite{santos2017thermodynamic} in the former, or the critical micelle concentration (cmc) in the latter, relatively regular aggregates spontaneously assemble. This transformation, however, is not a phase transition but rather a crossover.
Quantitatively locating of the ccd (or the cmc) is thus observable dependent, and determining the optimal approach is left to some degree of interpretation~\cite{johnston2016toward,santos2017thermodynamic}.
A canonical simulation approach for detecting the cmc involves identifying a marked bend in the pressure equation of state~\cite{amos1998osmotic,floriano1999micellization}. Recent simulations, however, suggest that this signature can sometime go missing for the clustering of systems with SALR interaction~\cite{santos2017thermodynamic}.

Second, the assembly dynamics of disordered microphases can be quite sluggish~\cite{schmalian2000stripe,geissler2004nature}. In experimental microphase formers, dynamically arrested amorphous gels and clusters are commonly observed~\cite{stradner2004equilibrium,campbell2005dynamical,klix2010structural,zhang2012non}.
Numerical simulations display remarkably slow equilibrium and out-of-equilibrium dynamics as well~\cite{tarzia2007lamellar,charbonneau2007phase,toledano2009colloidal,del2010microscopic,de2011dynamical}.
In addition, a recent numerical study suggests that the dynamics of disordered microphases is itself remarkably rich~\cite{zhuang2017communication}. Dynamical crossovers were found to accompany the clustering and percolation of both particles and voids. Sampling configurations of the disordered microphase regime is thus challenging, and no robust simulation approach has yet been formulated. Ad hoc mixtures of local and global particle displacements\cite{chen2000novel}, collective cluster moves\cite{whitelam2007avoiding}, and parallel tempering\cite{RN116,zhuang2016equilibrium} have been considered, but the relative merits of one or the other remain unclear. 

\begin{figure}[h]
 	\centering
 	\includegraphics[width = 0.45\textwidth]{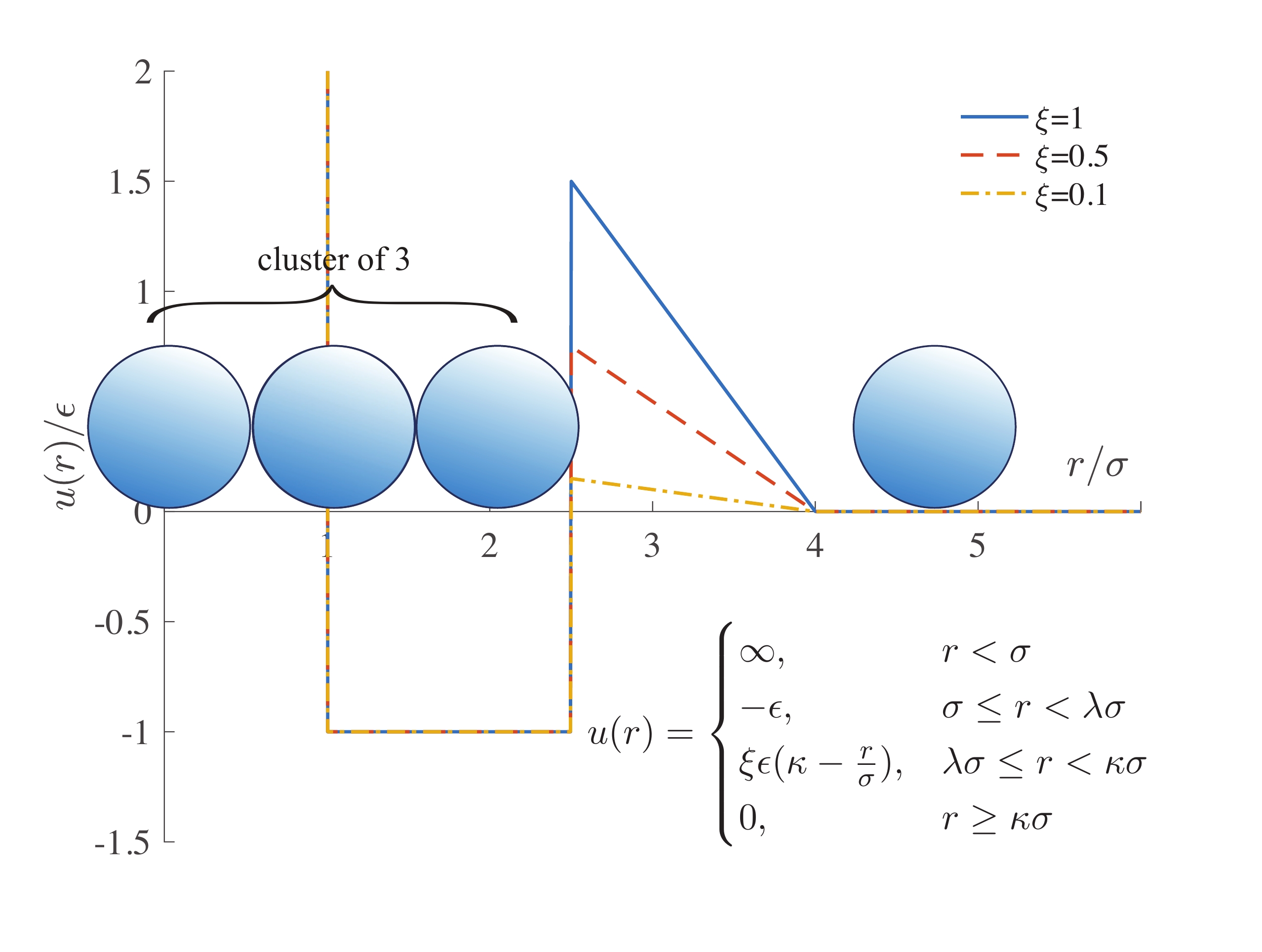}
 	\caption{The SWL model has a hard core repulsion for $r<\sigma$, followed by a square well attraction over $\sigma<r< \lambda \sigma$ and then by a linearly decaying repulsion for $\lambda \sigma<r< \kappa \sigma$.}
 	\label{fig:salrpot}
\end{figure}

In this work, we consider a one-dimensional (1D) archetype of SALR interactions. Although 1D models with finite-range interactions cannot undergo phase transitions\cite{van1950integrale,lieb2013mathematical}, they can nonetheless display a ccd\cite{pekalski2013periodic,pekalski2015bistability,ciach2017exactly}. In addition, the thermodynamics of such a model can be computed analytically by transfer matrices and its assembly dynamics can be straightforwardly simulated.  
We consider a square-well-linear (SWL) potential (Fig.~\ref{fig:salrpot}). This SALR interaction has a hard-core diameter, $\sigma$, that implicitly sets the unit of length, an attraction strength, $\epsilon$, that implicitly sets the unit of energy and is felt up to $\lambda \sigma$. Beyond this point a repulsive ramp of strength $\xi \epsilon$ decays linearly up to $\kappa \sigma$.
This particular potential form has been extensively studied in three dimensions~\cite{shen1993simple,sanz2009colloidal,cigala2015aggregate,RN116,zhuang2016equilibrium,zhuang2017communication},
and a closely related form was also considered in two dimensions\cite{Haw:2010}.
This interaction potential is known to display the same qualitative behavior as other SALR potentials, including those with a Yukawa repulsive form~\cite{hansen2000effective,lobaskin2001effective,schneider2004discontinuous,carlsson2003protein,footnote}. 

The rest of the paper is organized as follows. In Section \ref{sec:transfer}, we describe the transfer-matrix approach for deriving the thermodynamic properties of the system, including the equations of state, the cluster distribution function (CDF) and the gap distribution function (GDF). Section \ref{sec:mc} introduces the simulation approaches. The thermodynamic results are then presented and analyzed in Section \ref{sec:thermo}, while the relaxation dynamics is discussed in Section \ref{sec:dynamics}. A brief conclusion follows in Section \ref{sec:conclu}.

\section{Transfer-matrix Method}
\label{sec:transfer}

Continuous-space 1D models with finite-range interactions can be solved using transfer matrices~\cite{kofke1993hard,RN160,kardar2007statistical}. Changing variables from absolute particle positions, $x$, to relative distances between neighboring particles, $s$, indeed transforms the configurational part of the isothermal-isobaric partition function (with fixed number of particles $N$, pressure $p$ and inverse temperature $\beta$) into
\begin{align} \label{NPTgeneral}
	Z_G(p,\beta) &= \int_0^\infty \prod_{i=1}^N d s_i e^{-\beta u_i(s_i, s_{i+1}, ...) -\beta p s_i},
\end{align}
where $u_i$ is the sum of pairwise interactions between particle $i$ and subsequent particles-- $i+1,i+2\ldots$ -- along the chain. For interaction potentials with a hard core and a finite interaction range, this transformation is exact, because the maximal number of such interactions is finite. In the SWL model (Fig.~\ref{fig:salrpot}), for instance, up to $k = \lceil \kappa \rceil - 1$ nearest neighbors can interact at once. By analogy with the three-dimensional SWL models studied in Ref.~\citen{RN116} we set $\kappa=4$, hence particles can interact with up to their third nearest neighbors.
Under periodic boundary conditions, Eq.~\eqref{NPTgeneral} can then be written using the transfer matrix $\bm{M}$
\begin{equation} \label{tmat}
	Z_G = \mathbf{Tr}(M^N) = \Iambda_{\max}^N,
\end{equation}
where each entry of $\bm{M}$ contains the Boltzmann weight for a particular choice of $(s_i, s_{i+1}, ..., s_{i+k-1})$ along each row and $(s_{i+1}, s_{i+2}, ..., s_{i+k})$ along each column. Note that only when a row and a column have matching $s_{i+1}, ..., s_{i+k-1}$, is the entry nonzero, i.e.,
\begin{align}  \label{tmatentry}
	\bm{M}_{ab} =
	\begin{cases}
	e^{-\beta u_i(s_i, s_{i+1}, ..., s_{i+k}) - \beta p s_i} ds_i, \\
  \qquad \qquad \qquad \hfill s_{i+1},\ldots,s_{i+k-1} \text{ match},&\\
	0, \hfill \text{otherwise}.&
\end{cases}
\end{align}
In the isothermal-isobaric ensemble the Gibbs free energy is given by $G = -\frac{1}{\beta} \log Z_G$, hence the average number density, $\rho\equiv N/V$, follows as
\begin{equation} \begin{aligned} \label{tmatdensity}
	\rho^{-1} &=- \lim_{N\rightarrow\infty}\frac{1}{N}\left(\pdv{\log Z_N}{\beta p}\right)_{\beta,N}
	= -\left(\pdv{\log \Iambda_{\max}}{\beta p}\right)_{\beta} \\
  &= - \frac{q^{-1} (\partial{M}/\partial(\beta p))_\beta q}{q^{-1} q \Iambda_{\max}}
  = - \frac{q^{-1} M' q}{q^{-1} q \Iambda_{\max}},
\end{aligned} \end{equation}
where $M'$ is a tangent matrix of $M$ with entries $M'_{ab} = \left(\pdv{M_{ab}}{\beta p}\right)_\beta$; $\bm{q}$ and $\bm{q}^{-1}$ are the right and left eigenvectors for $\Iambda_\mathrm{max}$, respectively.

Formally, $M$ is an infinite matrix, but different discretization schemes can be used to reduce its size under a given numerical accuracy~\cite{godfrey2015understanding}.
The first and the most intuitive implementation is to use an $m$-part isometric discretization of $s_i$ over the interval $(1, \kappa)$.
Because particle overlaps, i.e., $s_i < 1$, are forbidden, the lower integration boundary is set to unity; because nearest neighbors don't interact if $s_i > \kappa$, then beyond that point only the ideal gas contribution,
\begin{equation}
	\bm{M}_{a(s_i>\kappa),b} = \int_{s_i=\kappa}^\infty e^{-\beta p s_i} d s_i =
	\frac{e^{-\beta p \kappa}}{\beta p},
	\label{eq:idgas}
\end{equation}
persists. We thus append this value to the end of the list of $s_i$.
This scheme results in a matrix that grows as $(m+1)^{k-1} \times (m+1)^{k-1}$, with $(m+1)^k$ nonzero entries. Although observables formally converge to their thermodynamic values as $m\rightarrow\infty$, the numerical accuracy at finite $m$ is affected by various aspects of the discretization scheme. For our discontinuous interaction potential, for instance, observables can oscillate with $m$.
Here, we choose a two-part discretization scheme that minimizes such error (See SI for details).
In the end, Matlab's iterative eigenvalue algorithms\cite{Matlab2017a} are used to obtain the matrix largest eigenvalues, $\Iambda_{\max}$, and corresponding eigenvector $\bm{q}$.

\subsection{Cluster Distribution Function Calculation (CDF)}
A 1D cluster is defined as a chain of $n$ particles with nearest-neighbor distances smaller than the SWL attraction range, i.e., $1 < s < \lambda$, with chain ends further than $\lambda$ away from the rest of the system. The CDF, $K(n) \equiv n r(n)$, is then the fraction of particles that belongs to a cluster of size $n$, with $r(n) \equiv \rho_\mathrm{n}(n)/\sum_{i=1}^\infty \rho_\mathrm{n}(i)$ being the fraction of clusters of size $n$.

The CDF can be computed from a transfer matrix scheme analogous to that used to measure spatial correlations in the 1D Ising model~\cite{RN169}. Because every entry of $\bm{M}$ (Eq.~\eqref{tmatentry}) corresponds to specific interparticle distances, $(s_i, s_{i+1}, ..., s_{i+k})$, a particle that is part of a cluster has null entries for $s_i > \lambda$. Using this \emph{masked} transfer matrix, $\bm{M}_\lambda$, the probability that any pair of neighboring particles belongs to a same cluster is
\begin{equation} \label{prob2}
\begin{aligned}
	P_\mathrm{clu}(2) &= \frac{\mathbf{Tr}(\bm{M}\bm{M_\lambda}\bm{M}...)}{\mathbf{Tr}(\bm{M}\bm{M}\bm{M}\bm{M}...)}
	= \frac{\mathbf{Tr}(\bm{M}_\lambda \bm{Q}\bm{D}^{N-1}\bm{Q}^{-1})}{\mathbf{Tr}(\bm{M}^N)} \\
	&= \frac{\bm{q}^{-1}\bm{M}_\lambda \bm{q}}{\bm{q}^{-1} \Iambda_{\max} \bm{q}},
\end{aligned}
\end{equation}
where the eigenvalue decomposition $\bm{M} = \bm{Q}\bm{D}\bm{Q}^{-1}$ simplifies the computation.
In general, the probability that $n$ particles belong to a cluster is thus
\begin{equation} \label{probn}
	P_\mathrm{clu}(n) = \frac{\bm{q}^{-1}\bm{M}_\lambda^n\bm{q}}{\bm{q}^{-1} \Iambda_{\max}^n \bm{q}}.
\end{equation}
The CDF can then be related to $P_\mathrm{clu}(n)$ as
\begin{align}
	1 = &P_\mathrm{clu}(1) = r(1)+2r(2)+3r(3) + ... \\
	    &P_\mathrm{clu}(2) = r(2) +2r(3) + ... \\
	    &P_\mathrm{clu}(3) = r(3) + ... \\
\Rightarrow
K(n) &= n r(n)  \\
&=n[P_\mathrm{clu}(n)+P_\mathrm{clu}(n+2)-2P_\mathrm{clu}(n+1)].\notag
\end{align}

\subsection{Gap Distribution Function (GDF)}
The distribution of gaps between neighboring particles~\cite{torquato1990nearest,torquato1995nearest}, can also be computed using transfer matrices
\begin{equation}
\begin{aligned}
	&P_{\mathrm{gap}}(s_i) = \frac{\bm{q}^{-1}\bm{M}_{(s_i)}\bm{q}}{\bm{q}^{-1} \Iambda_{\max} \bm{q}} 
	=  \frac{1}{\bm{q}^{-1} \Iambda_{\max} \bm{q}} \cdot \\
  &\sum_{(s_{i+1}, s_{i+2})} \bm{q}^{-1}(s_i, s_{i+1})\bm{M}{(s_{i}, s_{i+1}, s_{i+2})}\bm{q}(s_{i+1}, s_{i+2}),
\end{aligned}
\end{equation}
where $\bm{M}_{(s_i)}$ is a \emph{masked} transfer matrix with only nonzero entries for a given nearest-neighbor distance $s_i$.
From the cluster definition and the relationship between the total number of clusters and the number of unbound nearest neighbors~\cite{lee2017cluster}, the cumulative GDF can be related to the cluster density
\begin{equation} \label{Cbind}
\begin{aligned}
	\rho_{s} &= \rho \int_{\lambda}^{\infty} P_{\mathrm{gap}}(s_i) d s_i = \rho \left(1-\int_{0}^{\lambda} P_\mathrm{gap}(s_i) d s_i \right) \\
	&= \rho C_\mathrm{unbound} = \rho (1-C_\mathrm{bound}),
\end{aligned}
\end{equation}
where $C_\mathrm{unbound}$ is the cumulative GDF for $s_i \ge \lambda$, i.e., the fraction of unbound nearest neighbors, and $C_\mathrm{bound} \equiv 1-C_\mathrm{unbound}$ is the fraction of bound nearest neighbors. 

\section{Monte Carlo Simulations\label{sec:mc}}
\begin{figure}
 	\centering
	\subfloat{\subfigimg[width = 0.24\textwidth]{a}{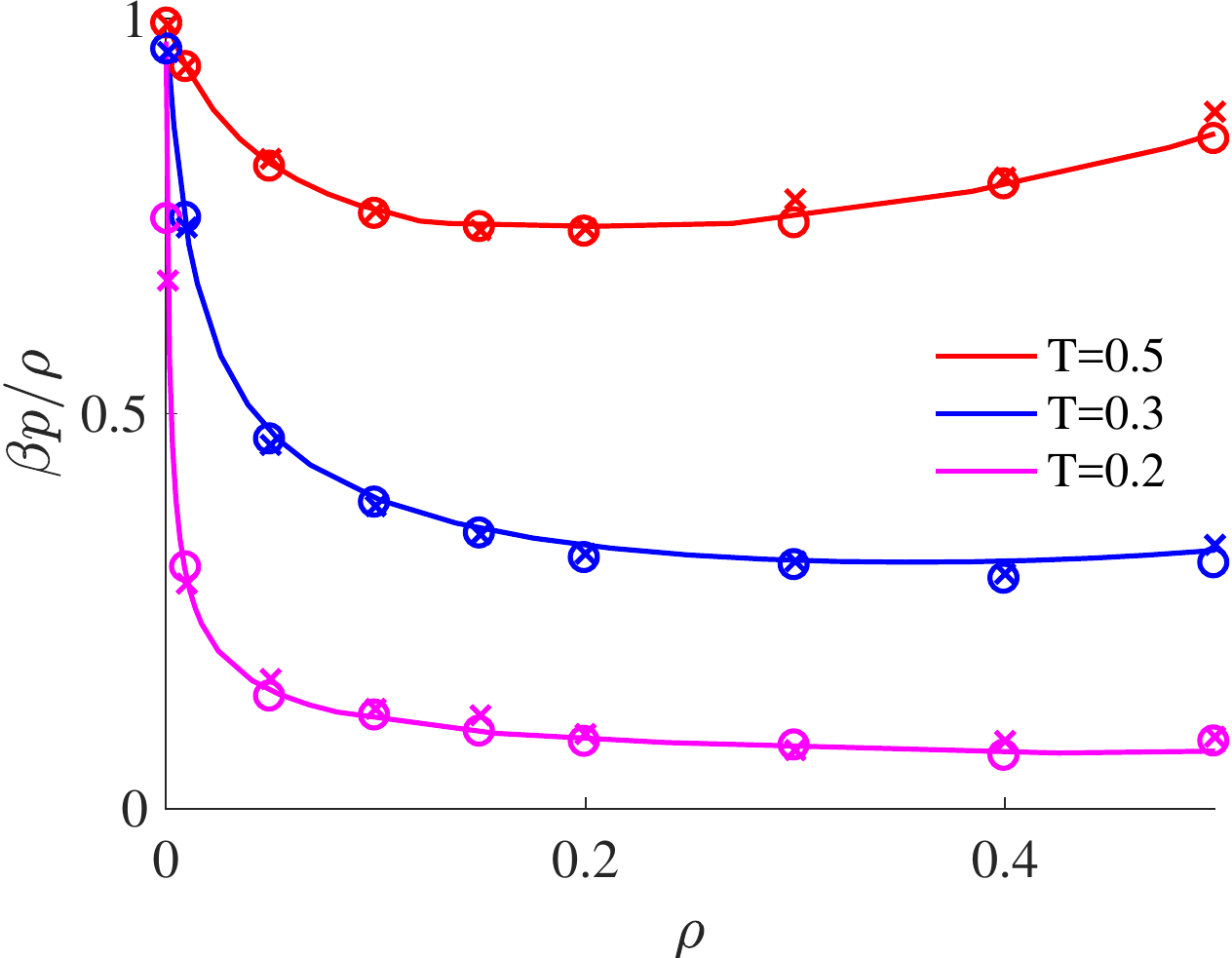}}
	\subfloat{\subfigimg[width = 0.24\textwidth]{b}{{cdf-T0.2-compare-eps-converted-to}.pdf}} \\
 	\caption{Comparison between the transfer-matrix calculations (lines), and the MMC (circles) and CHMC (crosses) simulation results for SWL with $(\lambda, \kappa,\xi) =
(2.5, 4, 1)$. (a) Equations of state at $T=0.5, 0.3$, and $0.2$. (b) CDF at $T=0.2$ for $\rho=10^{-4},\, 10^{-3},\, 10^{-2} \text{ and } 10^{-1}$.}
 	\label{algocomp}
\end{figure}

This section describes the Monte Carlo simulation algorithms used to study the equilibration of the SWL model: conventional Metropolis Monte Carlo (MMC), Heatbath Monte Carlo (HMC), their cluster-move versions (CMMC and CHMC), as well as Event-Chain Monte Carlo (ECMC). Note that the numerical implementation was validated by comparing simulation and transfer-matrix results (Fig.~\ref{algocomp}). Unless otherwise specified, all simulations are run for systems with $N=1000$ particles under a periodic boundary condition.

\subsection{Metropolis and Heatbath Monte Carlo \label{subsec:MMC}}

The main purpose of considering MMC here is 
as a generic local algorithm, with which to compare the more elaborate schemes described below. Our MMC implementation follows that of Kapfer et al~\cite{kapfer2017irreversible}. More specifically, the trial move for a particle i, is $\tilde{x}_i = x_i \pm \gamma \ell_{\mathrm{free}}$, where $\ell_\mathrm{free} = V/N-1$ is the average free volume per particle, and $\gamma$ is a random number uniformly distributed within $[0, 1)$. If a trial move changes the particle order, i.e. if $\tilde{x}_i < x_{i-1}$ or $\tilde{x}_i > x_{i+1}$, the move is automatically rejected. Otherwise, the move is accepted using the standard Metropolis criterion
\begin{equation}
\frac{\text{acc}(o \rightarrow n)}{\text{acc}(n \rightarrow o)} = \max[1,\exp\{-\beta[U(n)-U(o)]\}],
\end{equation}
where $U(n)$ and $U(o)$ are the total potential energy of the new  and old configuration, respectively.

HMC is a variant of MMC specifically tailored for 1D systems~\cite{kapfer2017irreversible}. Because SWL particles cannot interpenetrate, trial moves are conducted with uniform probability within the space confined by two nearest neighbors, and then accepted using the standard Metropolis criterion. 

In both cases, the pressure is then determined from the virial
\begin{align}
\label{virial}
	&\frac{\beta p-\rho}{\rho^2}= - \beta \int r g(r) \dv{u(r)}{r}dr \notag\\
  &= \int r g(r) e^{\beta u(r)} \dv{e^{-\beta u(r)}}{r} dr \\
 	&= \sigma g(\sigma^{+}) + \lambda \sigma [g(\lambda \sigma^{+}) - g(\lambda \sigma^{-})] + \frac{\beta \xi \epsilon}{\sigma} \int_{\lambda \sigma^+}^{\kappa \sigma} r g(r) dr,\notag
\end{align}
where the first term accounts for the hard-core repulsion, the second for the discontinuity between the attractive and the repulsive regimes and the third for the linear repulsion. The first two are evaluated by extrapolating the radial distribution function, $g(r)$, from the different sides of the discontinuity, denoted ``$+$'' and ``$-$'', while the third is obtained by standard numerical integration.

For both MMC and HMC, the simulation time, $t$, is computed in units of Monte Carlo sweeps, which include $N$ trial displacements. Note that this definition differs from that of Ref.~\citen{kapfer2017irreversible} by a factor of $N$.

\subsection{Cluster Monte Carlo}
At low temperatures, interparticle attraction results in a high rejection rate of attempted single-particle moves.
The spontaneous formation of aggregates, however, suggests that cluster displacements might then facilitate sampling. In order to preserve microscopic reversibility, clusters are here identified probabilistically. Pairs of sufficiently close neighbors are linked with probability $P_\mathrm{link}$.
In order to impose detailed balance, a trial displacement of the cluster move is attempted, and the new configuration is accepted with probability
\begin{equation} \label{clustermc}
\begin{aligned}
\frac{\text{acc}(o \rightarrow n)}{\text{acc}(n \rightarrow o)} &= \min[1,\exp\{-\beta[U(n)-U(o)]\} \cdot \\
&\prod_{ij} \frac{1-P_\mathrm{link}^r(i,j)}{1-P_\mathrm{link}^f(i,j)}],
\end{aligned}
\end{equation}
where $i$ denotes particles within the cluster and $j$ the others. Superscripts $f$ and $r$ denote the forward and reverse moves, respectively~\cite{frenkel2001understanding}.
Following the SWL model structure, the probability of linking a particle pair is chosen to depend on the attraction strength at that distance\cite{whitelam2007avoiding}
\begin{equation}
P_\mathrm{link}(i, i')\equiv
\begin{cases}
1-e^{-\beta \epsilon},& r_{ii'} < \lambda \\
0,& r_{ii'} \geq \lambda
\end{cases}.
\end{equation}
Clusters are then treated as quasi-particles with their collective displacements akin to those of single particles. Because $P_\mathrm{link}$ thus vanishes at high temperatures, hence the algorithm reduces back to single-particle MC scheme.

This clustering scheme is applied to both MMC and HMC, thus giving rise to Cluster Metropolis Monte Carlo (CMMC) and the Cluster Heatbath Monte Carlo (CHMC), respectively. In CHMC, for instance, a cluster trial displacement is conducted with uniform probability within the space contained by its two nearest neighbors and then accepted using the criterion given in Eq.~\eqref{clustermc}. For the purpose of comparing algorithmic dynamics, one attempted cluster displacement of $n$ particles is deemed equivalent to $n$ attempted MMC or HMC displacements.

\subsection{Event-chain Monte Carlo (ECMC)}

\begin{figure}
 	\centering
 	\includegraphics[width = 0.25\textwidth]{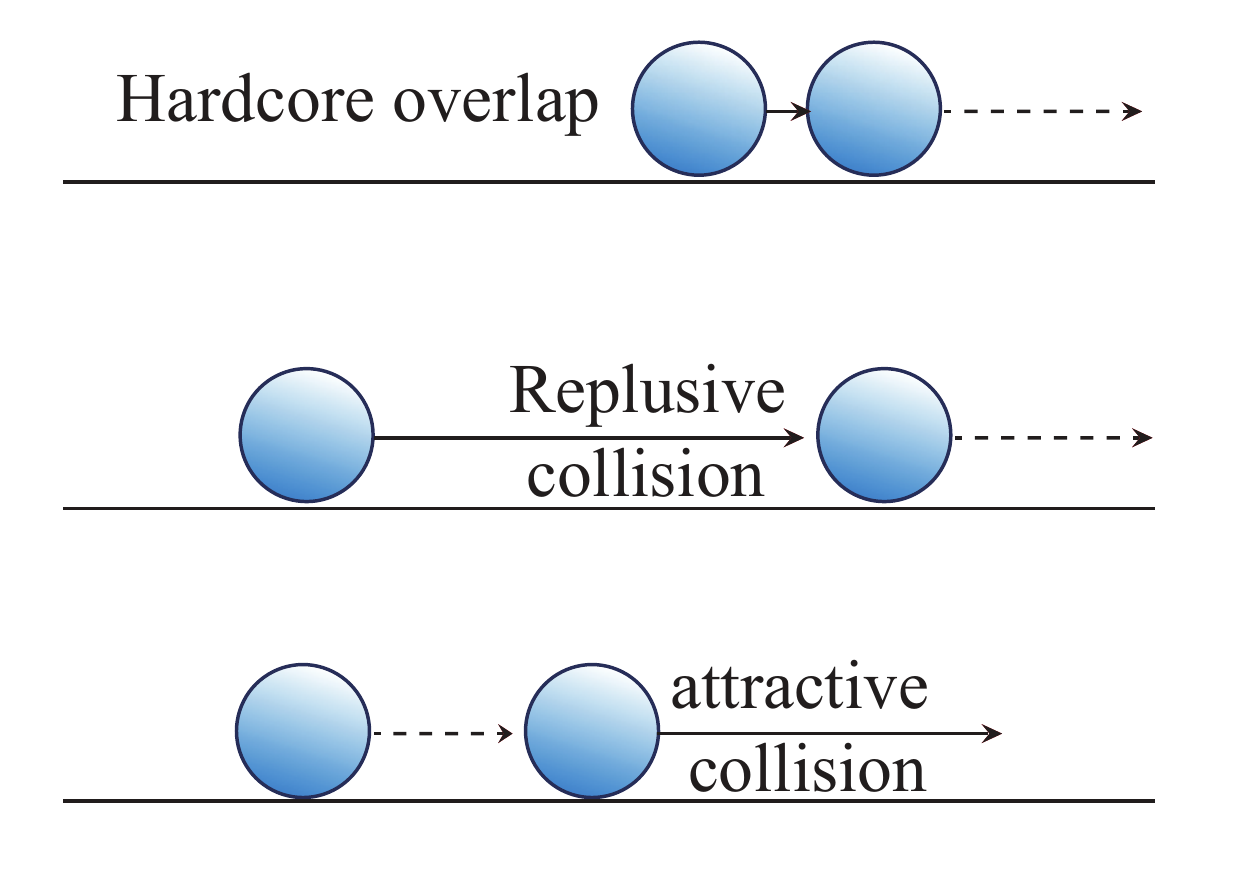}
 	\caption{Three types of ECMC collisions are possible for the SWL model: hard core; from non-interacting to repulsing; leaving the attractive well, from top to bottom. In the first two particles get closer, while in the third they moves apart.}
 	\label{fig:ecmcdemo}
\end{figure}

ECMC uses an altogether different strategy to accelerate phase space sampling~\cite{michel2014generalized}.
Unlike traditional MC, which first generates a trial move and then accepts it probabilistically, ECMC introduces sequential moves that resemble momentum transfer in Newtonian dynamics. The diffusion-like behavior of the Markov chains generated by MMC and its variants is thus vastly superseded by the ECMC dynamics~\cite{bernard2009event, michel2014generalized}.
ECMC notably balances realistic dynamics with moves that help escape spatial traps in systems of dense hard disks~\cite{bernard2011two}. For SALR potentials and other clustering models, however, the performance of ECMC (and related algorithms) has not been specifically examined. 

ECMC first assigns a moving direction, $\bm{e}$, to particle $i$, and then generates an admissible potential increase $E_{ij}^*=- \log \gamma_{ij}$ with respect to neighboring particle $j$, where $\gamma_{ij}$ is a random variable flatly distributed between $[0, 1)$. This random variable selects how far particle $i$ moves, $y_{ij}$, before a \emph{collision} happens. The collision location of particle $i$ is determined from the increase of its interaction with $j$,
\begin{equation}
	E_{ij}^* = \int_0^{E_{ij}^*} d [E_{ij}]^+ = \int_0^{y_{ij}} \left[ \pdv{E_{ij}(\bm{r}_j - \bm{r}_i - s)}{s} \right]^+ ds.
\label{ecmcscheme}
\end{equation}
If the energy increase is never higher than $E_{ij}^*$, no real solution for $y_{ij}$ exists and thus no collision between $i$ and $j$ takes place. The collision distance between these two particles is then set to infinity. The actual displacement of $i$ is the shortest $y_{ij}$ to all other particles, i.e., $y_i = \min_j(\{y_{ij}\})$, and the particle to subsequently move is $\argmin_j (y_{ij})$. For the SWL model, three collision types are possible (See Fig.~\ref{fig:ecmcdemo}).
The event chain terminates when the summation over all particle displacements equates a preassigned value, $l$. It is here chosen to be $0.1 N \ell_{\mathrm{free}} = 0.1 (V-N)$.

Interestingly, the system pressure can be determined directly for ECMC sampling -- thus sidestepping the virial in Eq.~\eqref{virial} -- by averaging over all event chains
\begin{equation}
	\frac{\beta p}{\rho} = 1 + \angles{\frac{\sum_{(i,j)}(x_j - x_i)}{l} }_{\text{event-chains}},
\end{equation}
where $(x_j - x_i)$ is the distance between two particles \emph{at the collision point}. It is negative for collisions that move particles apart, and positive otherwise. For the purpose of comparing algorithms, one collision event is deemed equivalent to one MMC or HMC attempted move, even though an ECMC collision is about twice as computationally demanding as a single MMC trial move~\cite{michel2014generalized}.

\section{Clustering Thermodynamics\label{sec:thermo}}
As mentioned in the introduction, even though long-range periodicity is not possible for a 1D SWL model, a crossover directly from a gas of particles to a gas of clusters ($\rho_\mathrm{ccd}$) can be observed at sufficiently low temperature. (At low density, there exists a corresponding critical clustering temperature, $T_\mathrm{cct}$.) Because the onset of clustering is but a crossover, its precise location is partly a matter of definition. In this section, we examine various schemes based on the equation of state and the CDF that have been proposed to identify the onset clustering in SALR systems.

\subsection{Equation of State \label{sc:state}}
\begin{figure}
 	\centering
 	\includegraphics[width = 0.48\textwidth]{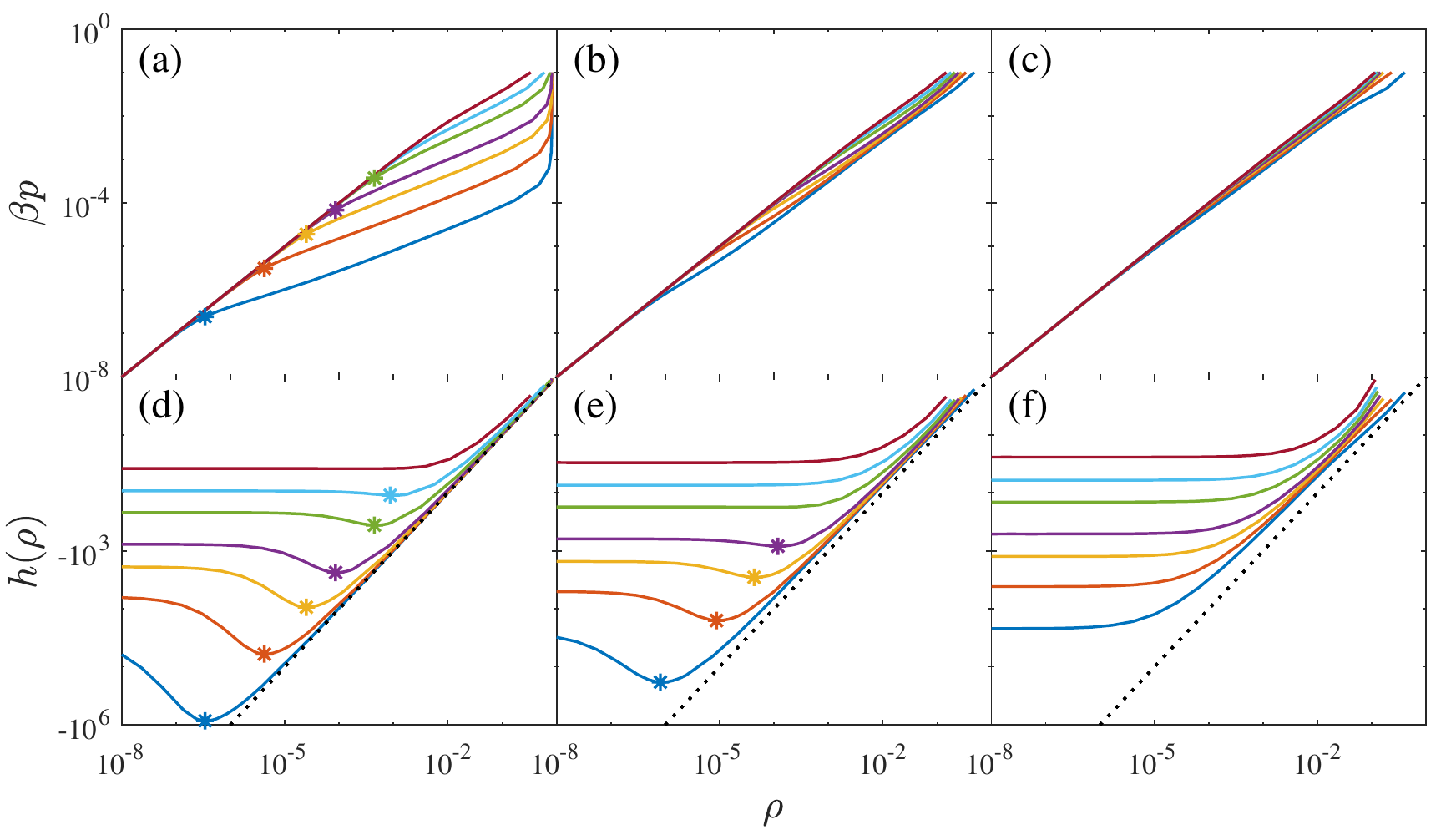}
 	\caption{Transfer-matrix results for the pressure, $\beta p$, and $h(\rho;T)$ for (a, d) $\lambda=2.5$, (b, e) $\lambda=2.2$, and (c, f) $\lambda=2.0$, in an SWL model with $(\kappa, \xi) = (4, 1)$ at $T=0.1, 0.12, 0.14, 0.16, 0.20, 0.24,$ and $0.30$, from bottom to top. Crossovers in $h(\rho;T)$ (asterisks) can be observed for $\lambda>2$ at sufficiently low temperatures. Changes to the equations of state are sufficiently pronounced to allow the identification of the crossover in (a) but not in (b). The scheme based on $h(\rho)$, by contrast, still detects the clustering crossover in (e). Note that neither approach detects a crossover for $\lambda \le 2$. Black dotted lines denote the cluster scaling, $~\rho^{-1}$, as described in text.}
 	\label{hfunc}
\end{figure}
A drastic change to the aggregation behavior of a system is expected to leave a trace on its macroscopic properties, such as its pressure.
The simplest proposed observable of this type derives from the long-established methods used for studying the cmc~\cite{amos1998osmotic,floriano1999micellization}. Typically, a micelle forming system transforms abruptly from a nearly ideal gas of particles to a nearly ideal gas of micelles. It was thus suggested that the point of largest curvature in the isothermal equation of state, $\beta p$, should be used to identify the ccd~\cite{santos2017thermodynamic}.

Because systems with SALR interactions can display large deviations from ideality both in the single-particle and in the cluster regimes, however, a direct application of this approach does not always clearly identify the ccd.
It was thus proposed that one should instead consider
\begin{equation}
	 h(\rho;T) = \frac{\beta p - \rho}{\rho^2}  = B_2(T) + B_3(T) \rho + \mathcal{O}(\rho^2),
	 \label{eqhrho}
\end{equation}
which specifically measures the deviation of the equations of state from ideality~\cite{RN116}. As can be seen from the corresponding expansion in terms of virial coefficients, $B_\alpha(T)$, if $B_3(T)$ is negative then $h(\rho;T)$ displays a minimum at a finite density. This minimum captures the onset of clustering.

The above two approaches are compared for various SWL models in Fig.~\ref{hfunc}. At low temperatures, the two schemes coincide (See panel a, d). At higher temperatures or smaller $\lambda$, changes to the equations of state, however, become fainter. Pinpointing a crossover then becomes markedly more arduous than in $h(\rho;T)$. A similar ambiguity was also reported in Ref.~\citen{santos2017thermodynamic}, where it was noted that the equation of state may not noticeably respond to clustering. The approach based on $h(\rho;T)$ thus appears slightly more robust.

Like the equation of state, $h(\rho;T)$ also provides microscopic insights into the clustering process. At densities above the crossover, but far from the harshly repulsive regime, we find that $h(\rho;T)\sim \rho^{-1}$ for all $T$. In this regime, the system thus effectively behaves as an ideal gas of clusters, with $\beta p \approx \rho/\hat{n}$ for an average cluster size $\hat{n}>1$, and hence $h(\rho) \approx -(1-1/\hat{n})/\rho$. This linear regime is especially clear at low temperatures for $\lambda > 2$ (Fig.~\ref{hfunc}), where trimers ($\hat{n}=3$) form preferentially. 
Note that for systems with no detectable crossover in $h(\rho;T)$ a linear regime can also be observed, but it is less clear and its intercept suggests that $\hat{n} \lesssim 2$. 

\subsection{Cluster Distribution \label{sec:CDF}}
\begin{figure}
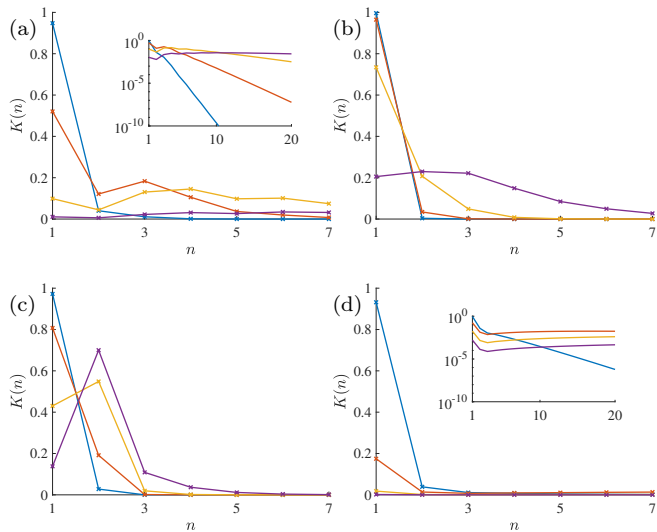

 	\centering
 	\subfloat{\subfigimg[width = 0.24\textwidth]{a}{{cdf-T0.2-eps-converted-to}.pdf}}
 	\subfloat{\subfigimg[width = 0.24\textwidth]{b}{{cdf-T0.4-eps-converted-to}.pdf}}\\
 	\subfloat{\subfigimg[width = 0.24\textwidth]{c}{{cdf-T0.2-lambda2-eps-converted-to}.pdf}}
	\subfloat{\subfigimg[width = 0.24\textwidth]{d}{{cdf-T0.2-xi0.1-eps-converted-to}.pdf}}\\
 	\caption{The CDF obtained from transfer matrices at $\rho=10^{-4}$ (blue), $10^{-3}$ (red), $10^{-2}$ (yellow) and $10^{-1}$ (purple), for $(\lambda, \xi)=(2.5, 1)$ at (a) $T=0.2$ and (b) $T=0.4$, (c) for $(\lambda, \xi)=(2, 1)$ at $T=0.2$, and (d) for $(\lambda, \xi)=(2.5, 0.1)$ at $T=0.2$. Insets in (a) and (d) present the same data on a lin--log plot.}
 	\label{cdf}
\end{figure}
We now consider a scheme that detects the onset of clustering directly from the CDF. According to the criterion proposed in Ref.~\citen{santos2017thermodynamic}, a separate peak in the CDF can sometimes be observed even in the absence of thermodynamic signatures. This is especially likely for small clusters.
Here, the limited typical cluster sizes in 1D system enable a finer assessment of the situation. 

Figure~\ref{cdf} presents the CDF for systems in different clustering regimes. Panel (a) shows that a separate peak at $n=3$ appears in the CDF between $\rho=10^{-4}$ and $10^{-3}$, which is consistent with $\rho_\mathrm{ccd}=4.7 \times 10^{-4}$. By contrast, panel (b) shows that no separate peak at $n=3$ appears when temperature is high, while panel (c) shows that for $\lambda = 2$ a peak at $n=2$ (instead of at $n=3$) emerges. Monomers then become dimers with no hint of additional clustering. We already know that this process is smooth and gradual, and it leaves no thermodynamic signature in $h(T;\rho)$. Hence, even though clustering does take place, no detectable ccd ensues. The distinction between dimer and trimer formation is thus reminiscent of that between submicellar clusters and micelles in higher-dimensional systems~\cite{johnston2016toward}. 

Panel (d) shows that at low frustration $\xi$ the size distribution of aggregates is much broader (see inset). In this case, assembly is the 1D echo of condensation, which is also a crossover (in absence of thermodynamic phase transitions) but is qualitatively distinct from the ccd. Although condensation and clustering leave similar signatures to the 1D equations of state, they affect the CDF quite differently. For the sake of comparison, we locate the condensation-like aggregation at the density where $K(3)=K(4)$. (In higher-dimensional systems, condensation is a first-order phase transition, with sharp features that easily distinguish it from ccd.) As proposed by Ref.~\citen{pekalski2013periodic}, condensation is here also accompanied by a crossover in the growth of the correlation length. By contrast to lattice models, however, trimer clustering leaves no such signature (See SI for details).

In summary, as density increases from the gas regime, in which the CDF monotonically decreases with increasing cluster size, three different clustering types can be distinguished: (i) condensation-like aggregation, marked by $K(3)=K(4)$; (ii) trimer formation (ccd), marked by $K(2)=K(3)$; and (iii) dimer formation, marked by $K(1)=K(2)$.

\subsection{Terminal Clustering Temperatures}

\begin{figure}[h]
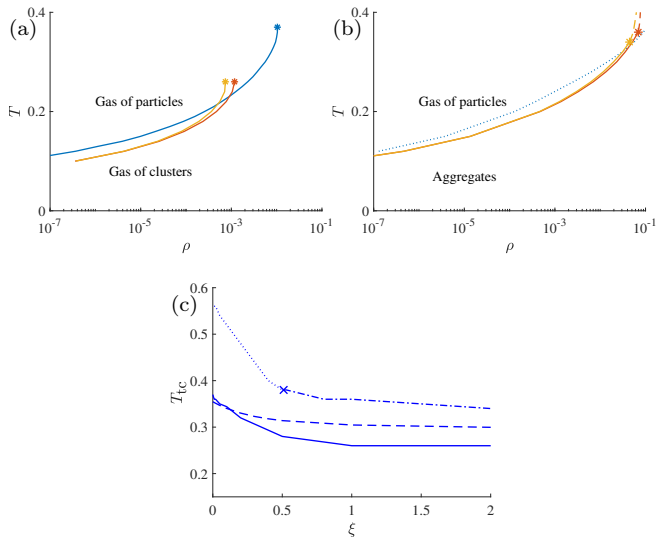

  \centering
  \subfloat{\subfigimg[width = 0.24\textwidth]{a}{{rhoc-2.5-eps-converted-to}.pdf}}
  \subfloat{\subfigimg[width = 0.24\textwidth]{b}{{rhoccdf-2.5-eps-converted-to}.pdf}} \\
  \subfloat{\subfigimg[width = 0.24\textwidth]{c}{{Tccdf-l2.5-eps-converted-to}.pdf}}
  \caption{(a) Temperature evolution of the onset of clustering extracted from $h(\rho; T)$ (as in Fig.~\ref{hfunc})
  for systems with different repulsion strengths, $\xi=0$ (blue), $1$ (red) and $2$ (yellow).
  For later reference, note that clustering takes place at $T<T_\mathrm{cct}=0.26$ for $\rho=0.001$.
  $T_{\mathrm{tc-}h}$ (asterisks) is estimated as the disappearance of the local minimum of $h(\rho;T)$.
  (b) Clustering crossovers determined from the CDF under the same conditions.
  Three types of clustering are then observed: (i) condensation-like aggregation (dotted line); (ii) particles to trimers (solid lines); and (iii) particles to dimers (dashed lines).
  $T_{\mathrm{tc-CDF}}$ (asterisks) is estimated as described in text.
  (c) $T_\mathrm{tc}$ estimated from $h(\rho;T)$ ($T_{\mathrm{tc-}h}$, solid line), $B_3=0$ ($T_{\mathrm{tc-}B_3}$, dash line) and the CDF ($T_{\mathrm{tc-CDF}}$). A dotted line identifies the clustering as being of type (i) to type (iii), and a dash-dotted as being of type (ii) to type (iii). The cross denotes the $\lambda$ transition from condensation to trimer clustering (See in text).}
  \label{crossT}
\end{figure}

Clustering disappears with increasing temperature, thus defining a terminal clustering $T_{\mathrm{tc}-X}$, where the subscript $X$ denotes the observable from which $T_\mathrm{tc}$ estimated.
The disappearance of the local minimum of $h(\rho;T)$ thus defines, $T_{\mathrm{tc}-h}$. Both trimer clustering and condensation-like aggregation also only exist for $T < T_\mathrm{tc-CDF}$. For $T>T_\mathrm{tc-CDF}$ dimers first form as density increases. The onsets of clustering given by these two measurements, when it exists, qualitatively agree with one another, as shown in Figure~\ref{crossT}a and b.

From Eq.~\eqref{eqhrho}, we further have that a minimum in $h(\rho)$ can only be observed if $B_3(T)<0$ (See SI). The sign change of $B_3(T)$ with temperature for $\lambda>2$ thus also provides an estimate, $T_{\mathrm{tc}-B_3}$ (Fig.~\ref{crossT}c). For $\lambda<2$, however, next-nearest neighbor interactions cannot be attractive, hence $B_3(T)\geq0,\,\forall T$. This explains why no crossover in Fig.~\ref{hfunc}f is ever observed. Whatever clustering might take place in this case leaves no thermodynamic signature. Note that for the limit case, $\lambda=2$, we have $T_{\mathrm{tc}-B_3}=0$.  
Note also that in the large $\xi$ limit $T_\mathrm{tc}$ in all cases tend a constant. Although in that case the thermodynamic cluster distribution can be sampled by transfer matrices, it would then be impossible for an unbound particle to cross the repulsive barrier using local dynamics, such as MMC.

The change from condensation-like to trimer clustering is controlled by $\xi$ (Fig.~\ref{crossT}c). In higher-dimensional or mean-field systems this transition is known as the $\lambda$ transition\cite{khanna1972bose,archer2007phase,kormann2014lambda}. Using the above definitions from one process and the other, we obtain $\xi_{\lambda}=0.51$ for $\lambda=2.5$ and $\xi_{\lambda}=0.46$ for $\lambda=2.2$.
In short, while Ref.~\citen{santos2017thermodynamic} found that clustering may or may not be accompanied by a detectable change to equation of state in 3D, the presence of a crossover in $h(\rho;T)$ in 1D seems tightly controlled by the size of the clusters (dimers vs trimers) that assemble.

\section{Relaxation Dynamics \label{sec:dynamics}}
Having identified the regime in which thermodynamic clustering takes place, we can examine the clustering dynamics using various MC algorithms. Two different initial conditions are considered:
(i) all particles forming a compact chain, as in Ref.~\citen{kapfer2017irreversible}; and 
(ii) equally-spaced particles, which is akin to instantaneously quenching a typical high-temperature configuration.
We then examine the relaxation to equilibrium of the fraction of bound nearest neighbors, $C_\mathrm{bound}$, which is related to the total cluster density (Eq.~\eqref{Cbind}).
Note that other preparations could be devised and that the equilibrium mixing time could also be considered, but their results are expected to be qualitatively similar to what is measured here.

\begin{figure}[h]
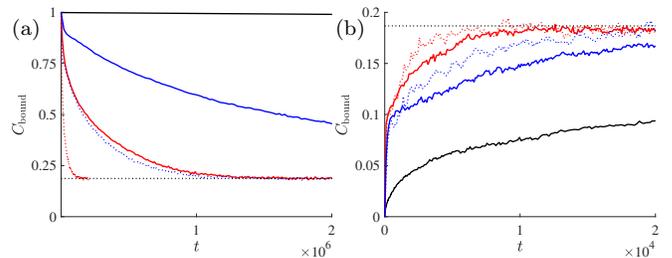

 	\centering
 	\subfloat{\subfigimg[width = 0.24\textwidth]{a}{{draw-T0.22,D0.001-eps-converted-to}.pdf}}
 	\subfloat{\subfigimg[width = 0.24\textwidth]{b}{{draw-eqdis-T0.22,D0.001-eps-converted-to}.pdf}} \\
 	\caption{The relaxation of $C_\mathrm{bound}$ to equilibrium (black dotted line) from (a) a compact chain and (b) equally-spaced particles, at $T=0.22$ and $\rho=10^{-3}$, averaged over 20 runs, using MMC (blue solid line), CMMC (blue dotted line), HMC (red solid line), CHMC (red dotted line), and ECMC (black solid line).}
 	\label{relaxraw}
\end{figure}

Figure~\ref{relaxraw} depicts the equilibration dynamics from both initial conditions for a state point that falls within the trimer clustering regime (cf.~Fig.~\ref{cdf}a)).
Similar to the micelle formation dynamics\cite{aniansson1974kinetics,johnston2016toward}, the structure of $C_\mathrm{bound}$ relaxes in two steps. The first corresponds to fast single-particle exchanges on a scale $\tau_1$, and the second to slow cluster turnovers on a scale $\tau_2$. Because $\tau_1 \ll \tau_2$, the relaxation (mixing) time $\tau_\mathrm{mix} \approx \tau_2$. For $t \gg \tau_1$, the decay is indeed exponential, which suggests the following long-time fitting form
\begin{equation}
	C_\mathrm{bound} = C_\mathrm{bound,eq}- C_0 e^{-t/\tau_\mathrm{mix}},
	\label{taumixdef}
\end{equation}
where $C_\mathrm{bound,eq}$ is the equilibrium result, and $C_0$ is a fitting constant. 

\begin{figure}[h]
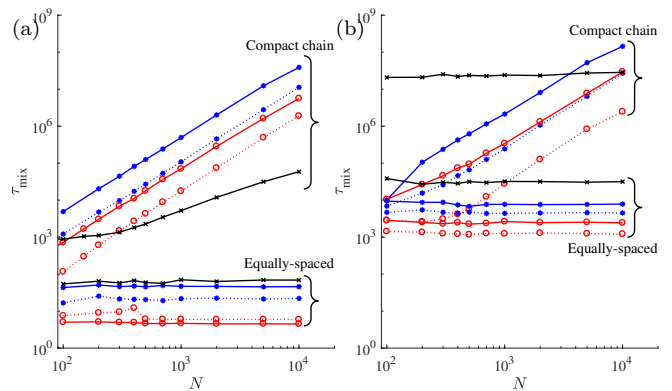

 	\centering
 	\subfloat{\subfigimg[width = 0.24\textwidth]{a}{{relax-T0.5,D0.1-eps-converted-to}.pdf}}
 	\subfloat{\subfigimg[width = 0.24\textwidth]{b}{{relax-T0.22,D0.001-eps-converted-to}.pdf}} \\
 	\caption{Evolution of the relaxation time, $\tau_\mathrm{mix}$, at (a) $\rho=0.1$, $T=0.5$ and (b) $\rho=0.001$, $T=0.22$ for MMC (blue solid line with asterisks), CMMC (blue dotted line with asterisks), HMC (red solid line with circles), CHMC (red dotted line with circles), and ECMC (black solid line with crosses). The relaxation time of compact chains increases with system size, but that of equally-spaced particles is independent.}
 	\label{relaxN}
\end{figure}

Reference~\citen{kapfer2017irreversible} showed that the relaxation of a compact chain of hard spheres with both MMC and HMC has an algorithmic complexity of $\tau_\mathrm{mix} \sim \mathcal{O}(N^2 \log N)$, while ECMC has $\mathcal{O}(N \log N)$. For a system with the SWL interaction, however, the scaling relation is more complex (See Fig.~\ref{relaxN}a). As expected, at high temperatures ECMC has a linear time complexity, while both single-particle and cluster-move MC scale as $\mathcal{O}(N^2)$ (including a logarithmic correction fits the data equally well). However, in the low-temperature clustering regime while the results of the cluster and single-particle algorithms still scale as $\mathcal{O}(N^2)$, the mixing time of ECMC remains nearly constant (Fig.~\ref{relaxN}b). 
The linear scaling comes with such a high prefactor that ECMC is far from optimal in this size regime. 

\begin{figure}[h]
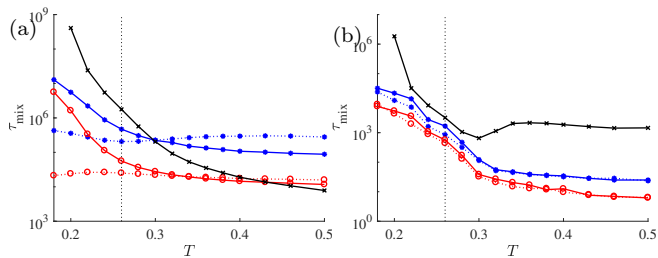

 	\centering
 	\subfloat{\subfigimg[width = 0.24\textwidth]{a}{{relax-T,D0.001-eps-converted-to}.pdf}}
 	\subfloat{\subfigimg[width = 0.24\textwidth]{b}{{relax-eqdis-T,D0.001-eps-converted-to}.pdf}} \\
 	\caption{Temperature evolution of the relaxation time, $\tau_\mathrm{mix}$, from (a) a compact chain and (b) an equally-spaced configuration at density $\rho=0.001$. Colors and symbols as in Fig.~\ref{relaxN}. For this density, Fig.~\ref{crossT}a gives $T_\mathrm{cct}=0.26$ (vertical dotted line). Note the difference in scale between (a) and (b).
  Data for $T \ge 0.24$ in (b) are obtained from $N=10000$ particles to ensure the numerical accuracy for small $C_\mathrm{bound,eq}$ cases.
  }
 	\label{relaxT}
\end{figure}

For a compact chain, the extent of the slowdown depends sensitively on the MC scheme. Single-particle algorithms experience a marked slowdown for $T<T_\mathrm{cct}$, but cluster algorithms show nothing comparable (Fig.~\ref{relaxT}a), which is consistent with the dynamical observation of Ref.~\citen{zhuang2017communication}.
As expected, for high temperature hard-sphere like system~\cite{kapfer2017irreversible}, ECMC has the shortest relaxation time, but at low temperatures CHMC wins the palm.
For this initial condition, cluster cleaving reaches the equilibrium CDF more efficiently than single-particle moves, which must produce a series of monomer exchanges to achieve a comparable result. Heatbath moves further accelerate sampling avoiding overlaps and efficiently (yet unphysically) surmounting the barriers to rearrangement.
In higher dimensional systems, global trial moves of this sort are not easily designed, but simpler versions have also been used to accelerate equilibration.
In Ref.~\citen{RN116}, for instance, aggregation-volume bias Monte Carlo\cite{chen2000novel} are used to enable surface-to-surface particle moves that sidestep the repulsion barrier.

For equally-spaced particles, the relaxation time is generally several orders of magnitude shorter than for the compact chain.
Because $\tau_{mix}$ remains constant with system size (Fig.~\ref{relaxN}), we conclude that equilibration is controlled by local processes. Because limited cluster cleavage is needed, heatbath and collective moves are here less significant, hence the relaxation dynamics of the advanced algorithms is within one order of magnitude of MMC even at low temperatures (Fig.~\ref{relaxT}b). CHMC is nonetheless still the fastest, although only by a small margin.

At low temperatures, ECMC dynamics gets increasingly sluggish for both initial conditions (Fig.~\ref{relaxT}). The onset of slowdown roughly coincides with $T_\mathrm{cct}$, but not as closely as for single-particle moves. The nature of the the event-chain algorithm underlies this effect. Because the admissible climb in energy at each step is set by the Boltzmann weight of barrier~\cite{michel2014generalized}, the probability of leaving the attraction range is $e^{-\beta(\xi(\kappa-\lambda)+1)}$, which for for Fig.~\ref{relaxN}b gives $\approx 1 \times 10^{-5}$. Hence, because bond breaking is rare, the relaxation time is large, e.g., $\tau_\mathrm{mix} \approx 10^7$ at $T=0.22$. 
As a result ECMC then mainly translates the system, which is a rather ineffective path to thermalization.
For a similar reason, at high temperatures ECMC relaxes equally-spaced particles more than an order of magnitude slower than the other schemes. (Strangely, $\tau_{mix}$ increases with temperature for over a brief temperature interval, $0.3 \le T \le 0.36$.)
In short, while ECMC is effective to equilibrate a compact chain through rapid long-range transport, it is far from optimal when equilibration mostly entail local processes.

\section{Conclusions \label{sec:conclu}}
By studying the thermodynamics and dynamics of a 1D SALR model, we have clarified the clustering assembly behavior of microphase formers. The model simplicity enabled the consideration of various proposals for detecting the onset of clustering and for evaluating the efficiency of different sampling algorithms. In both cases, these observations translate into a clearer grasp of the algorithmic ambiguities previously encountered. In particular, we conclude that the function $h(\rho; T)$ characterizes the pressure response to the onset of clustering more finely than earlier macroscopic approaches.
We have also determined the extent to which different types of collective Monte Carlo moves can accelerate sampling in these systems.
We expect these insights to apply to a broad range of SALR models, and thus inform subsequent simulation efforts.

\section*{Acknowledgments}
We thank J.~P\c{e}kalski and Y.~Zhuang for stimulating discussions. We acknowledge support from the National Science Foundation's Research Triangle Materials Research Science and Engineering Center (MRSEC) under Grant No. (DMR-1121107) and from NSF's grant from the Nanomanufacturing Program (CMMI-1363483).
\beginsupplement
\begin{widetext}

\section*{Supplementary information}

\section{Discretization choice in Transfer-matrix method}

In the main text transfer matrices for the isothermal-isobaric ensemble are used to compute the system density and various other observables. For a fixed discretization scheme, the numerical error of this approach grows as temperature decreases or density increases. To analyze the main contributions to this error, we here investigate the results of computations under extreme conditions (for this article), $T=0.2$ and $p=0.01$ for $(\lambda, \kappa, \xi)=(2.5, 4, 1)$. Note that in this case, the thermodynamic density is $\rho=0.60148(1)$.

\subsection{Isometric Discretization}

\begin{figure}[h]
 	\centering
 	\includegraphics[width = 0.5\textwidth]{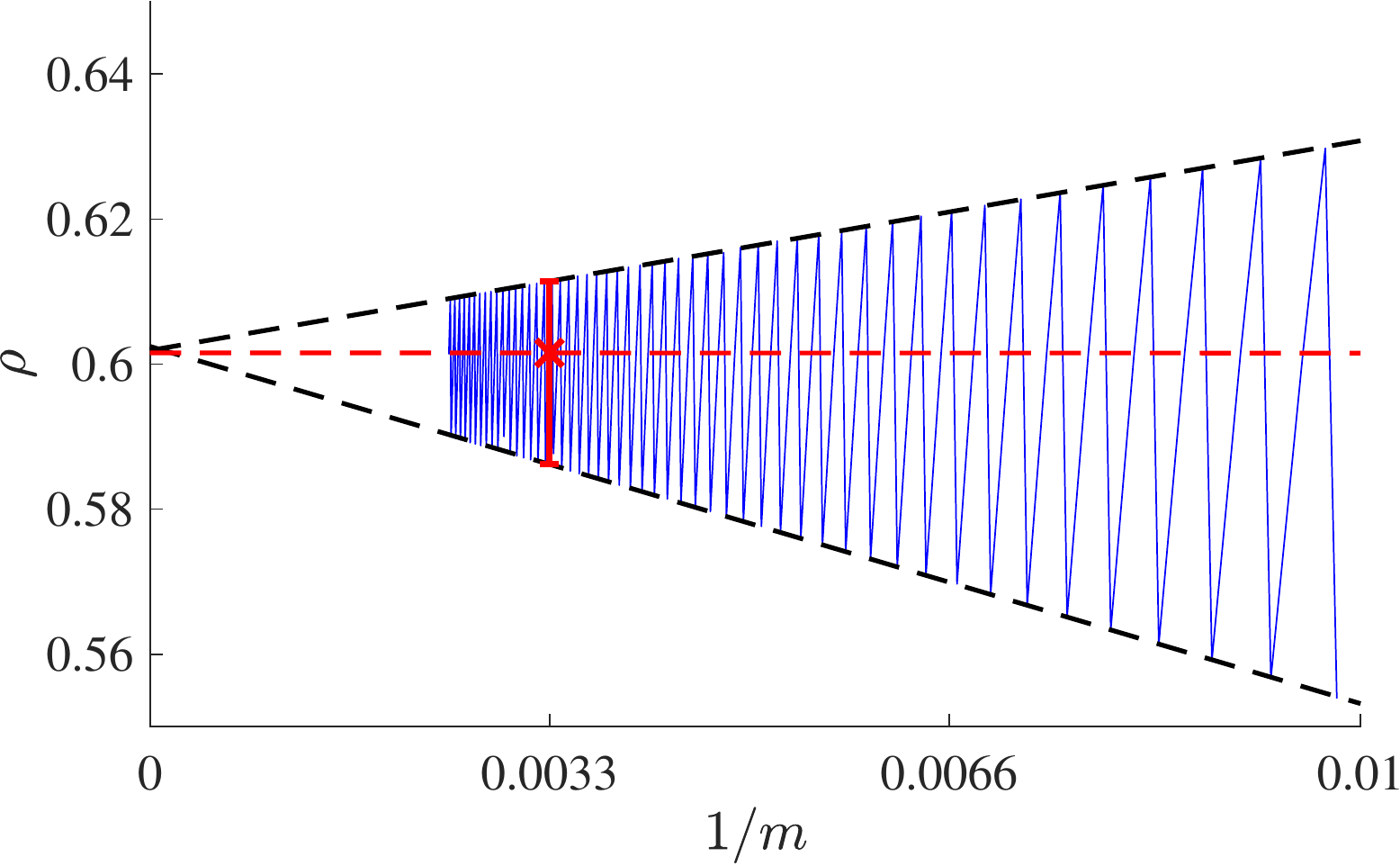}
 	\caption{Convergence of the isometric discretization scheme with $m$. The asymptotic convergence in $1/m$ is accompanied by oscillations. Around $m=300$, oscillations range $\sim3\%$, but thanks to a fortuitous cancellation choosing $(m \mod 6) = 3$ (red dash line) reduces this error to a fraction of a percent. For $m=303$, in particular, the error is $0.05\%$.
  }
 	\label{fig:oscil}
\end{figure}

We first consider an isometric discretization of $s_i \in (1, \kappa)$. Figure~\ref{fig:oscil} shows that density estimates for this scheme oscillate with discretization number $m$. The period corresponds to the number of bins needed to go from one integer subdivision of the attraction well to another. In the case of $(\lambda, \kappa)=(2.5, 4)$, $(m \mod 6)=0$ sets the lower limit of the oscillation and $(m \mod 6)=1$ it upper limit. The virial (main text, Eq.~\eqref{art-virial}) suggests that this behavior might be related to the two discontinuities in the interaction potential: $r=1$ and $r=\lambda$. Even though the transfer matrix includes third-nearest neighbor (3NN) interactions, these don't give rise to any discontinuity, hence only the NNN transfer matrix is here of interest.

\begin{figure}[h]
 	\centering
 	\includegraphics[width = 0.75\textwidth]{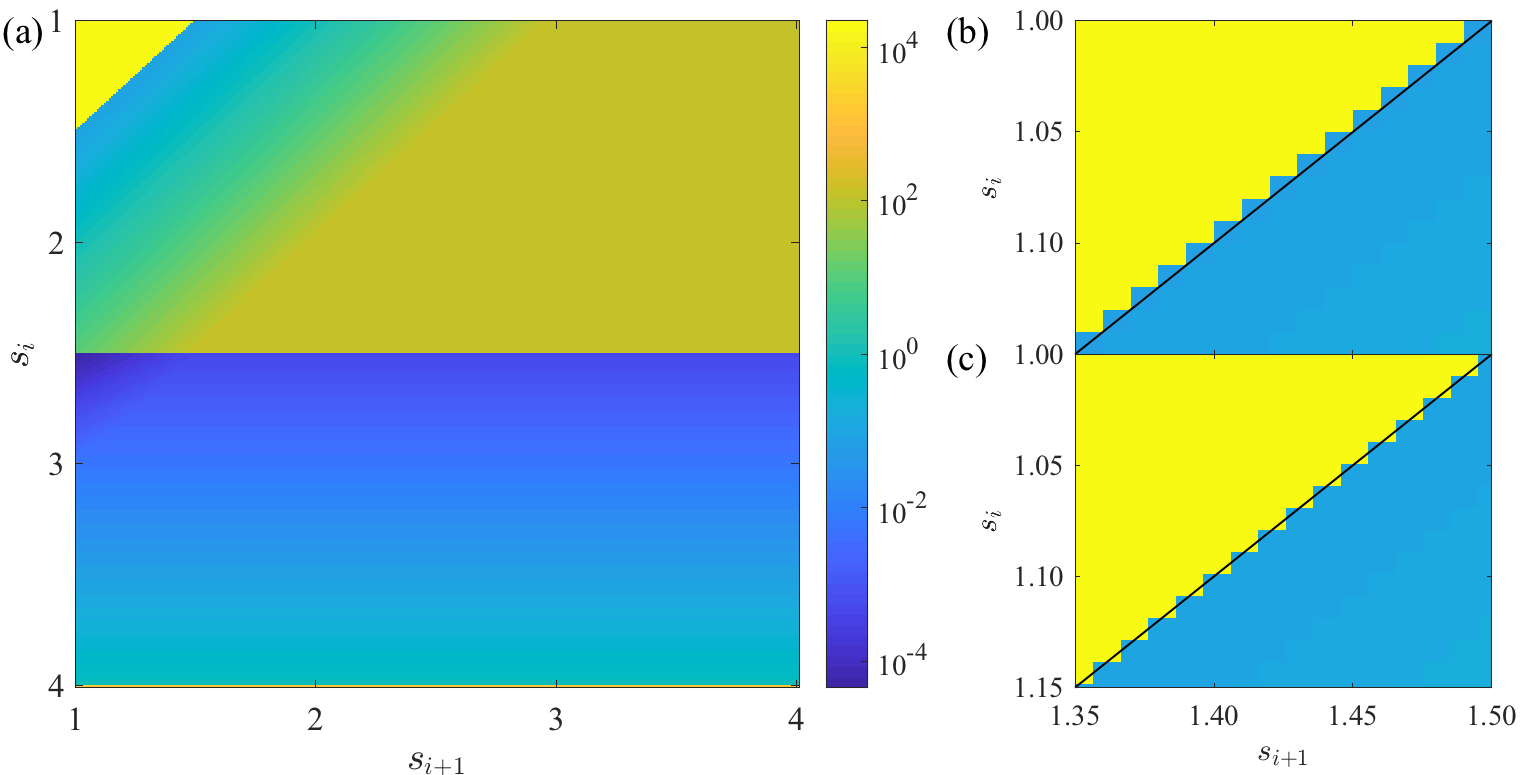}
 	\caption{Transfer matrix for NNN interactions.
 	(a) The whole matrix for $m=300$;
  (b) Detail of (a) around the potential discontinuity at $s_i+s_{i+1} = \lambda$ (black solid line);
  (c) Detail of the same area for $m=303$. Color encodes the magnitude of the entry on a logarithmic scale.}
 	\label{fig:tmatshow}
\end{figure}

Visualizing the transfer matrix helps identify the numerical origin of this oscillation (Fig.~\ref{fig:tmatshow}).
The main Boltzmann weights are found in the upper-left triangle of side $m/6$, which corresponds to next-nearest neighbor attraction regime. The hypotenuse coincides with the discontinuity of NNN interaction at $s_i+s_{i+1} = \lambda$. When $(m \mod 6)=0$, the hypotenuse
coincides with the right edge of the discontinuity for these entries (Fig.~\ref{fig:tmatshow}b). Because the fraction of entries that cross the discontinuity boundary is 
\begin{equation} \label{eq:errorpf}
    \frac{2m/6}{(m+1)^2} \sim \frac{1}{m},
\end{equation}
the error must asymptotically converge as $1/m$.  As can be seen in Fig.~\ref{fig:tmatshow}c, a fortuitous cancellation surprisingly takes place for $(m \mod 6)=3$. Optimizing this parameter is, however, not generally satisfying. More elaborate discretization schemes are necessary to reduce the error more systematically.


\subsection{Simpson's Rule}

\begin{figure}[h]
  \centering
  \subfloat{\subfigimg[width = 0.33\textwidth]{a}{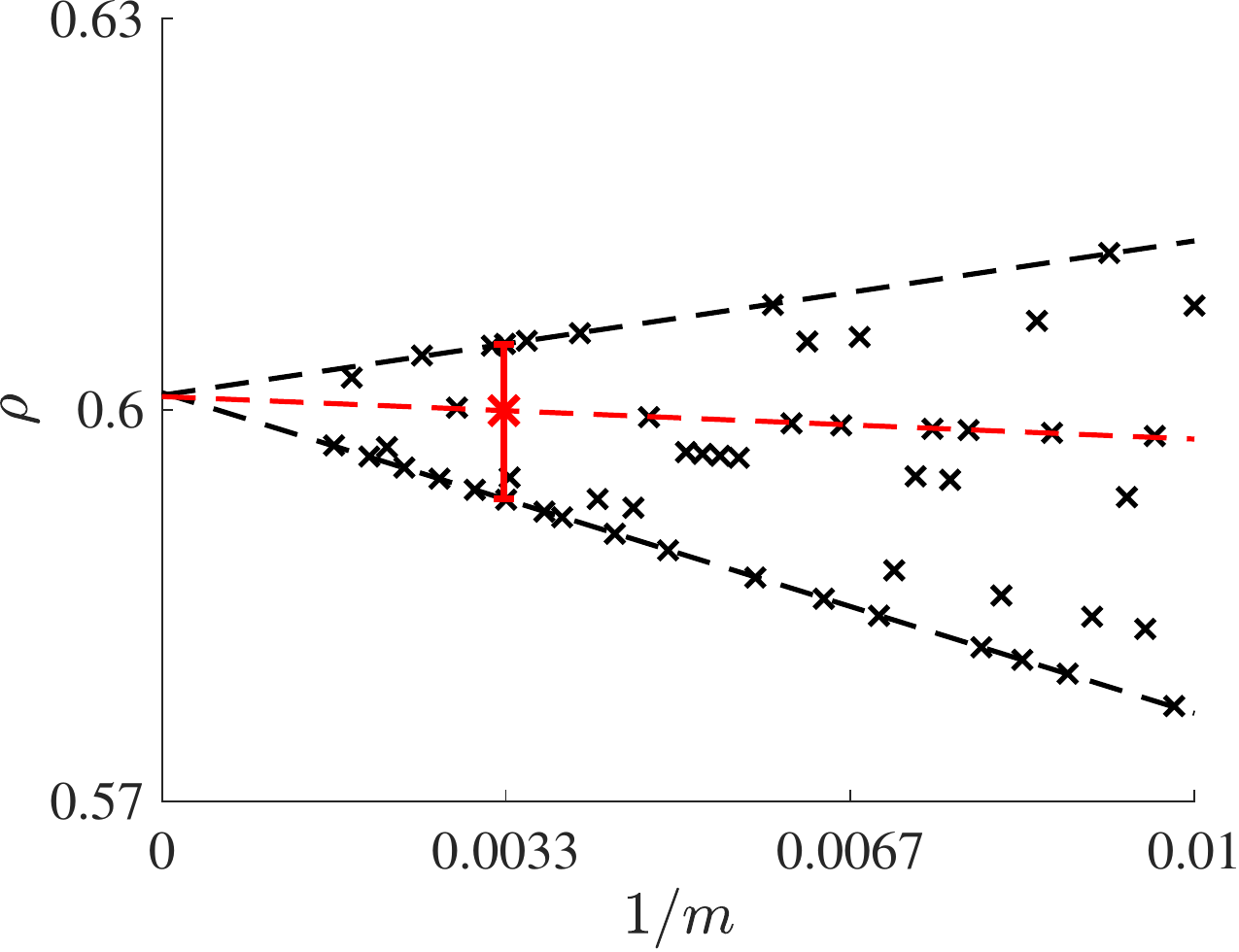}}
  \subfloat{\subfigimg[width = 0.33\textwidth]{b}{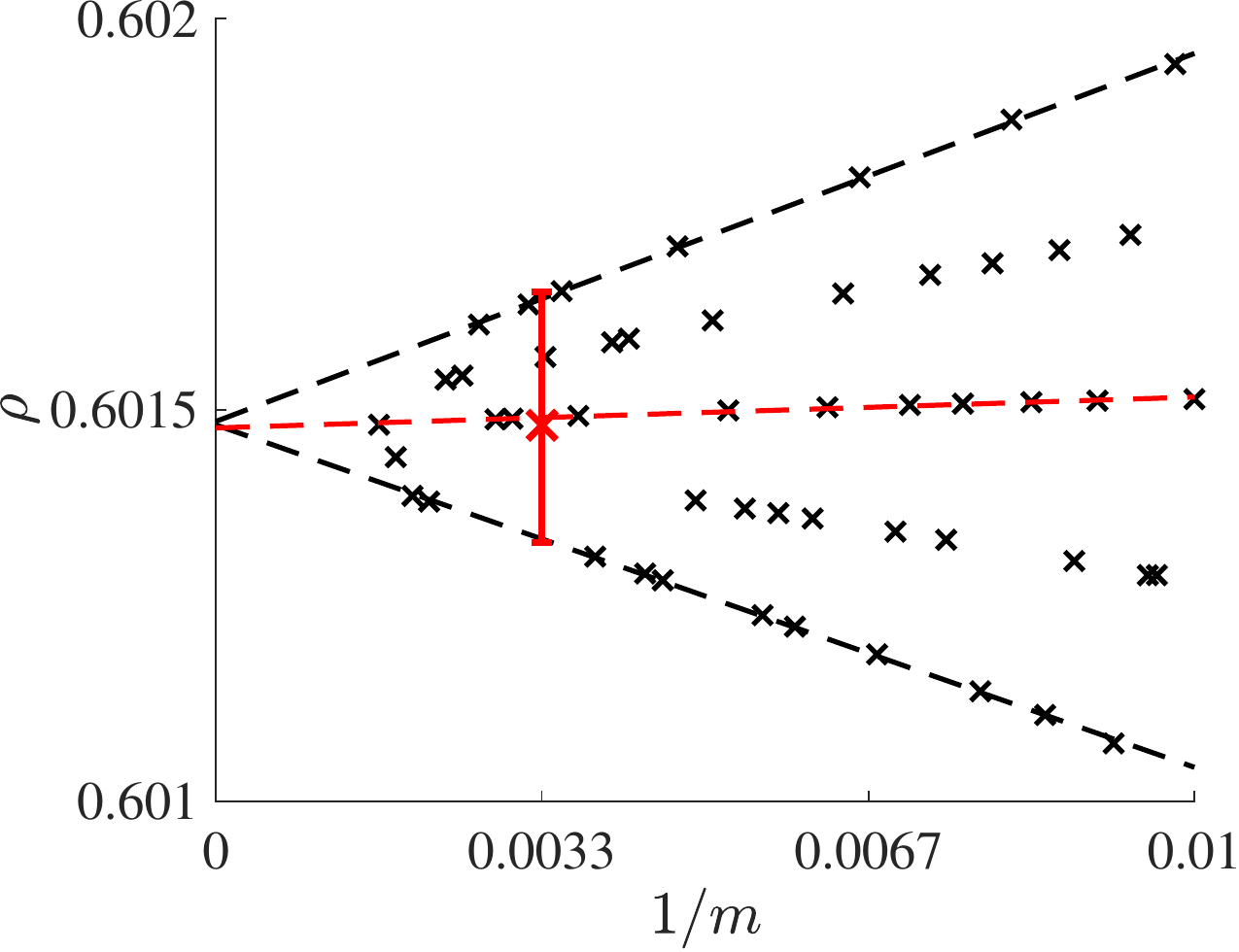}} \\
  \caption{Density for different discretization schemes: (a) Simpson's rule and (b) two-part discretization. The results still converge as $1/m$, but now with a much smaller prefactor.}
  \label{fig:alteralgos}
\end{figure}

To approximate the Boltzmann weight for an entry more precisely, an analogy to Simpson's rule for numerical integration is proposed:
\begin{equation}
\begin{aligned}
    M(s_i, s_{i+1}) &= \int_{s_i - \delta s/2}^{s_i+\delta s/2}  \int_{s_{i+1} - \delta s/2}^{s_{i+1}+\delta s/2}  e^{-\beta (u(s_1) + u(s_1+s_2) + p s_1)} d s_1 d s_2  \\
    &= \frac{1}{36} \left\{ e^{-\beta u(s_1- \delta s/2) -\beta p (s_1 - \delta s/2)} (e^{-\beta u(s_1+s_2 - \delta s)} + 4 e^{-\beta u(s_1+s_2 - \delta s/2)} + e^{-\beta u(s_1+s_2)}) \right.\\
    &+  4 e^{-\beta u(s_1) -\beta p s_1} (e^{-\beta u(s_1+s_2 - \delta s/2)} + 4 e^{-\beta u(s_1+s_2)} + e^{-\beta u(s_1+s_2+\delta s/2)}) \\
    &+  \left. e^{-\beta u(s_1 + \delta s/2) -\beta p (s_1 + \delta s/2)} (e^{-\beta u(s_1+s_2 - \delta s)} + 4 e^{-\beta u(s_1+s_2 - \delta s/2)} + e^{-\beta u(s_1+s_2)} ) \right\},
\end{aligned}
\end{equation}
where $\delta s_i = (\kappa-1)/m$ denotes the interval of discretization. This general approach halves the oscillation strength compared to the midpoint rule (Fig.~\ref{fig:alteralgos}a).

\subsection{Two-part Discretization}

Discretizing more finely the region of the matrix that contains the largest weights is also expected to improve the numerical accuracy. For the SWL model, this region corresponds to $s_i + s_{i+1} < \lambda$. For example, dividing the list of $s_i$ into two parts $s_\mathrm{A} \in (1, s_\mathrm{d})$ and $s_\mathrm{B} \in (s_\mathrm{d}, \kappa)$, with $m = m_\mathrm{A} + m_\mathrm{B}$ and isometric discretization intervals $\delta s_\mathrm{A} < \delta s_\mathrm{B}$. Under the midpoint sampling rule, the error resulting from the discontinuity at $s_i+s_{i+1} = \lambda$ is reduced if $m_\mathrm{A}$, $\delta s_\mathrm{A}$ and $s_\mathrm{d}$ satisfy
\begin{equation}
        s_\mathrm{d}-1  = m_\mathrm{A} \delta s_\mathrm{A} = \lambda-2+\delta s_\mathrm{A}/2.
\end{equation} 
This choice minimizes the error because the division coincides with the discontinuity boundary regardless of $m$, as for $(m \mod 6)=3$ in Fig.~\ref{fig:tmatshow}c. For $m=300$, this two-part discretization gives $\rho=0.60149$, which indistinguishable from the asymptotic value (Fig.~\ref{fig:alteralgos}b). 


This scheme was implemented for the various calculations in this article.  Because the choice $(T=0.2,\,p=0.01)$ is an extreme case, we conclude that the result reported in this article have at most $0.1\%$ error. This scale is smaller than the line width in the figures of the main text.


\section{Correlation Length}
The spatial correlation as a function of particle separation is defined as
\begin{equation}
  G(i, j) = \angles{(s_i -\angles{s_i})(s_j -\angles{s_j})} = \angles{s_i s_j} - \angles{s_i}\angles{s_j}
\end{equation}
Generalizing the derivation described in Ref.~\citen{varga2011structural}, it can be shown that the correlation decays exponentially when $\lvert i-j \rvert \rightarrow \infty$. The correlation length is then $\xi_\mathrm{L} = \log(\Iambda_\mathrm{max}/\lvert \Iambda_2 \rvert)^{-1}$, where $\Iambda_2$ is the second dominant eigenvalue of $M$.

\begin{figure}[h]
  \centering
  \includegraphics[width = 0.95\textwidth]{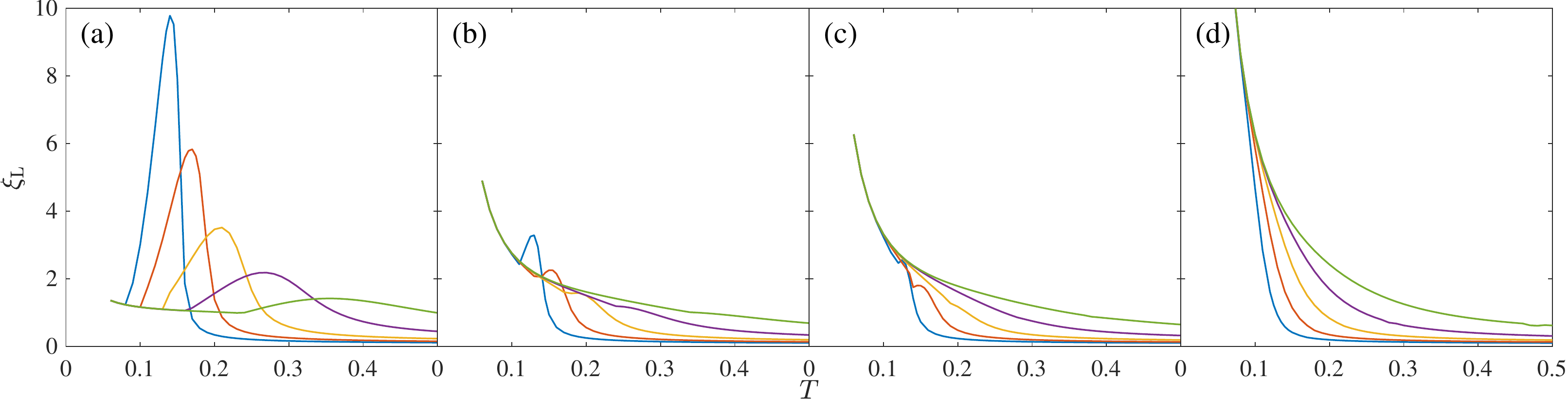}
  \caption{Correlation length, $\xi_\mathrm{L}$, for different repulsion strength, (a)$\xi=0.1$, (b)$0.5$, (c)$0.6$ and (d)$1$ for $\rho=10^{-5}$ (blue), $10^{-4}$ (red), $10^{-3}$ (yellow), $10^{-2}$ (purple) and $10^{-1}$ (green) . Note that the $\lambda$ transition identified by CDF is $\xi_\lambda=0.51$.}
  \label{fig:corlen}
\end{figure}

In 1D SALR lattice models, the correlation length was found to display a marked growth at the onset of clustering~\cite{pekalski2013periodic}.
Here, although the correlation length also grows with decreasing temperature at large $\xi$, its magnitude changes gradually and displays no remarkable feature around the onset of clustering. However, a separate peak does appear at small $\xi$, when the system undergoes condensation-like aggregation (Fig.~\ref{fig:corlen}). Here, the correlation length thus only captures ordering on length scales longer than that of the trimers. 

\section{Virial Coefficients Calculation}

The second and third virial coefficients are obtained by integrating the Mayer function $f(r)$: 
\begin{align}
  B_2(T) &= -\frac{1}{2} \int f(r)dr, \\
  B_3(T) &= -\frac{1}{3} \iint f(r)f(r')f(r-r') dr dr',
\end{align}
where
\begin{equation}
  f(r)=e^{-\beta u(r)}-1 = 
  \begin{cases}
    -1, & r<1, \\
    e^\beta-1, & 1 \le r < \lambda, \\
    e^{-\beta \xi (\kappa-r)}-1, & \lambda \le r < \kappa, \\
    0, & r \ge \kappa.
  \end{cases}
\end{equation}
The integral for $B_2(T)$ can be evaluated analytically
\begin{equation}
  B_2(T) = -e^{-\beta(1-\lambda)} + \frac{1-e^{-\beta\xi(\kappa-\lambda)}}{\beta\xi} + 2 \lambda - \kappa.
\end{equation}
The analytical form of $B_3(T)$ is, however, somewhat more involved. It is here obtained by numerical integration (Fig.~\ref{fig:B3}).

\begin{figure}[H]
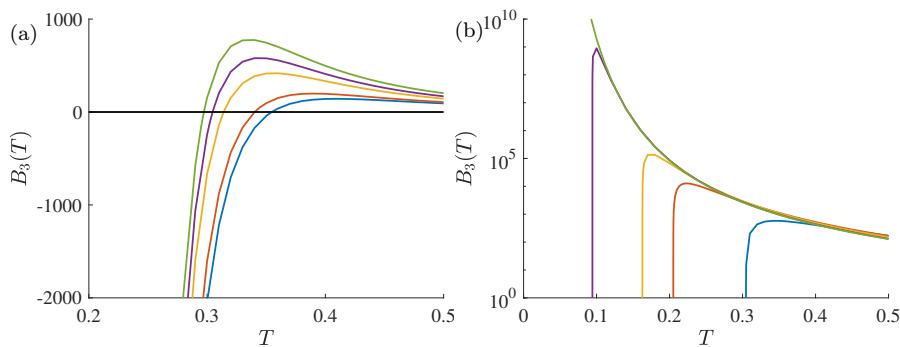

  \centering
  \subfloat{\subfigimg[width = 0.33\textwidth]{a}{{B3-2.5-v2-eps-converted-to}.pdf}}
  \subfloat{\subfigimg[width = 0.33\textwidth]{b}{{B3-xi1-eps-converted-to}.pdf}}
  \caption{Third virial coefficient $B_3(T)$ under (a) $\lambda=2.5$, where $\xi=0,0.1,0.5,1$ and $4$, from bottom to top; (b) $\xi=1$, where $\lambda=2.5, 2.2, 2.1, 2.01$ and $2.0$, from right to left.}
  \label{fig:B3}
\end{figure}
 
The terminal clustering temperature $T_\mathrm{tc}$ can be estimated by solving $B_3(T_\mathrm{tc})=0$. 
For the 1D SWL potential $\kappa\leq4$ this condition has to be strictly followed for clustering to be possible; in general higher-order coefficients can give rise to clustering even if $B_3(T)>0$.
Figure~\ref{fig:B3}a shows that the zero of $B_3(T)$, when it exists, decreases with $\xi$, as illustrated in Fig.~\ref{crossT}c of main text. This zero vanishes at $\lambda=2$ (Fig.~\ref{fig:B3}b).

\end{widetext}
\bibliography{abbrev} 
\end{document}